\RequirePackage{fix-cm}
\documentclass[twocolumn,epjc3-upd,fleqn]{svjour3}
\smartqed  
\RequirePackage{graphicx}
\RequirePackage{subfigure}
\usepackage{rotating}
\usepackage{amsmath}
\usepackage{mathtools}
\RequirePackage[colorlinks,citecolor=blue,urlcolor=blue,linkcolor=blue]{hyperref}
\usepackage{doi}
\usepackage[]{units}
\usepackage{xspace}
\usepackage[switch]{lineno} 
\usepackage{cite}
\journalname{Eur. Phys. J. C}
\def\qino{\mbox{CUORICINO}\xspace}

\def\tect{$^{130}$Te\xspace}

\def\teod{TeO$_2$\xspace}

\def\bbz{$0\nu\beta\beta$\xspace}

\def\vita-dim{$\tau_{1/2}$}

\def\be{\begin{equation}}
\def\ee{\end{equation}}

\begin{document}
\title{Double-beta decay of $^{130}$Te to the first $0^+$ excited state of $^{130}$Xe with CUORE-0}
\author{C.~Alduino\thanksref{USC} 
\and
K.~Alfonso\thanksref{UCLA} 
\and
D.~R.~Artusa\thanksref{USC,LNGS} 
\and
F.~T.~Avignone~III\thanksref{USC} 
\and
O.~Azzolini\thanksref{INFNLegnaro} 
\and
T.~I.~Banks\thanksref{BerkeleyPhys,LBNLNucSci} 
\and
G.~Bari\thanksref{INFNBologna} 
\and
J.~W.~Beeman\thanksref{LBNLMatSci} 
\and
F.~Bellini\thanksref{Roma,INFNRoma} 
\and
A.~Bersani\thanksref{INFNGenova}
% ...
} 
\institute{Department of Physics and Astronomy, University of South Carolina, Columbia, SC 29208 - USA\label{USC} 
\and
Department of Physics and Astronomy, University of California, Los Angeles, CA 90095 - USA\label{UCLA} 
\and
INFN - Laboratori Nazionali del Gran Sasso, Assergi (L'Aquila) I-67010 - Italy\label{LNGS} 
\and
INFN - Laboratori Nazionali di Legnaro, Legnaro (Padova) I-35020 - Italy\label{INFNLegnaro} 
\and
Department of Physics, University of California, Berkeley, CA 94720 - USA\label{BerkeleyPhys} 
\and
Nuclear Science Division, Lawrence Berkeley National Laboratory, Berkeley, CA 94720 - USA\label{LBNLNucSci} 
\and
INFN - Sezione di Bologna, Bologna I-40127 - Italy\label{INFNBologna} 
\and
Materials Science Division, Lawrence Berkeley National Laboratory, Berkeley, CA 94720 - USA\label{LBNLMatSci} 
\and
Dipartimento di Fisica, Sapienza Universit\`{a} di Roma, Roma I-00185 - Italy\label{Roma} 
\and
INFN - Sezione di Roma, Roma I-00185 - Italy\label{INFNRoma} 
\and
INFN - Sezione di Genova, Genova I-16146 - Italy\label{INFNGenova} 
% ...
}

\thankstext{e1}{E-mail: cuore-spokesperson@lngs.infn.it}
\thankstext{n1}{Deceased}
\date{Received: date / Accepted: date}
\twocolumn
\maketitle
\begin{abstract}

We report on a search for double beta decay of $^{130}$Te to the first $0^{+}$ excited state of $^{130}$Xe using a \unit[9.8]{kg$\cdot$yr} exposure of $^{130}$Te collected with the CUORE-0 experiment. In this work we exploit different topologies of coincident events to search for both the neutrinoless and two-neutrino double-decay modes. We find no evidence for either mode and place lower bounds on the half-lives: \unit[$\tau^{0\nu}_{0^+}>7.9\cdot 10^{23}$]{yr} and \unit[$\tau^{2\nu}_{0^+}>2.4\cdot 10^{23}$]{yr}.  Combining our results with those obtained by the CUORICINO experiment, we achieve the most stringent constraints available for these processes: \unit[$\tau^{0\nu}_{0^+}>1.4\cdot 10^{24}$]{yr} and \unit[$\tau^{2\nu}_{0^+}>2.5\cdot 10^{23}$]{yr}. 


\end{abstract}

\section{Introduction}
\label{sec:intro}
Two-neutrino (2$\nu\beta\beta$)~\cite{1} and neutrinoless (0$\nu\beta\beta$) ~\cite{2} double beta decay are among the rarest decay processes studied.
While the former is allowed by the Standard Model and has been experimentally detected in a number of isotopes~\cite{3}, the latter has never been observed; its discovery would imply that
lepton number is not conserved and that neutrinos are in fact Majorana particles~\cite{14,15}.


The CUORE experiment (Cryogenic Underground Observatory for Rare Events)~\cite{16,29,30,31}, which is currently running at Laboratori Nazionali del Gran Sasso (LNGS), is designed to perform a high-sensitivity search for $0\nu\beta\beta$ decay of $^{130}$Te to the ground state of $^{130}$Xe\cite{9}. The active isotope is contained in TeO$_{2}$ crystals, which are operated as thermal detectors in a cryostat capable of reaching temperatures below 10 mK. At this temperature, the crystal heat capacity becomes very small and consequently a release of energy within a crystal results in a detectable increase of its temperature. The sought-after experimental signature of 0$\nu\beta\beta$ decay is a monochromatic peak in the summed energy spectrum of the final state electrons at 2527.518 $\pm$ 0.013 keV~\cite{22,23,24}, which is the transition energy of the decay. To maximize the sensitivity of the search, the radioactive background at the transition energy must be kept as low as possible. 

\begin{figure}
\begin{center}
\includegraphics[width=\linewidth]{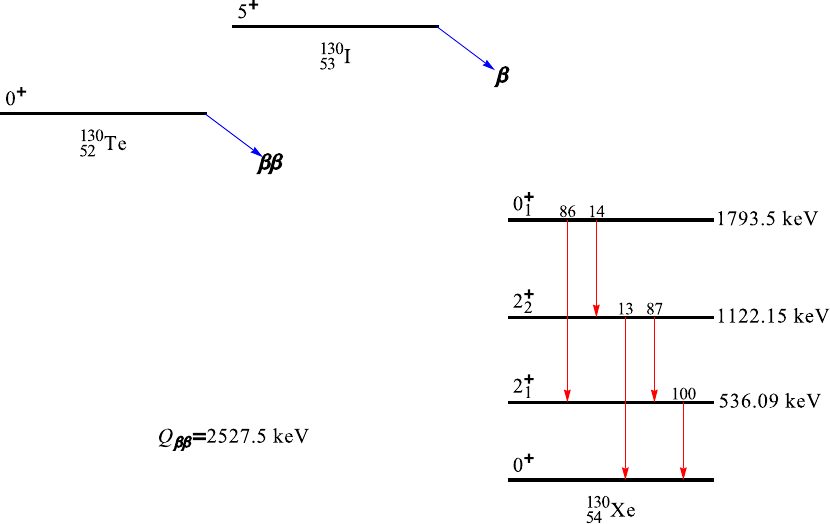} %
\caption{Decay scheme of $^{130}$Te showing the energy levels and the branching ratios for the $\gamma$ rays \cite{9}.}
\label{fig:decay_scheme}
\end{center}
\end{figure}

The first tower assembled in the CUORE assembly line was operated as a standalone experiment, named CUORE-0~\cite{5}, from 2013 to 2015. CUORE-0 was designed to validate several key aspects of CUORE, including detector construction, data acquisition and the analysis framework. In addition to this, CUORE-0 provided a sensitive probe of several rare decays, including {\bbz}\cite{20,8} and $2\nu\beta\beta$\cite{10} decay of ${}^{130}$Te to the ground state of ${}^{130}$Xe, and the $\beta^{+}EC$ decay of ${}^{120}$Te\cite{21}. In this work we focus on a search for double-beta decay of ${}^{130}$Te to the first $0^+$ excited state of ${}^{130}$Xe ($\beta\beta_{0^+}$) with CUORE-0. As shown in Figure \ref{fig:decay_scheme} this decay emits two electrons, which share a maximum energy of 734~keV, followed by a gamma cascade to the ground state of ${}^{130}$Xe. The most probable de-excitation pattern, which has a branching ratio of 86.0\%, involves the emission of two gamma rays with energies of 1257.4 keV and 536.1 keV. Two more patterns are also possible, namely the emission of three gamma rays with energies of 536.1 keV, 586.0 keV and 671.3 keV (branching ratio of 12.2\%) and the emission of two gamma rays with energies of 671.3 keV and 1122.2 keV (branching ratio of 1.8\%). These gamma lines result in multi-detector coincidence signatures which we exploit in our analysis to achieve very powerful background rejection.
As is the case for the decay to the ground state, the summed energy spectrum of the emitted electrons is distinctly different in the $2\nu\beta\beta_{0^+}$ vs.\ the $0\nu\beta\beta_{0^+}$ case. The former is a continuous spectrum ($0-734$ keV), whereas the latter is a monochromatic
peak centered at 734 keV.\\

Prior to the current work, the most stringent constraints on these decays came from the \qino~\cite{13} experiment, a predecessor to CUORE-0, which reported the following limits on the decay half-lives~\cite{12}: 
\begin{center}
$\tau_{0^+}^{0\nu}>9.4\cdot 10^{23}$ yr\,, \\
$\tau_{0^+}^{2\nu}>1.3\cdot 10^{23}$ yr
\end{center}

\section{Experiment}
\label{sec:exp}
The CUORE-0 tower, just like all the CUORE towers, contains 52 $^{\text{nat}}$TeO$_2$ crystals (i.e., $^{130}$Te is present at its natural abundance of 34.2\%\cite{25}). The crystals are arranged in a copper frame into 13 floors, with each floor containing four crystals~\cite{5}. The mass of each crystal is \unit[750]{g}, for a total detector mass of \teod of \unit[39]{kg} or \unit[10.8]{kg} of \tect.

The crystals are operated as thermal detectors at a working temperature of $\sim$\unit[10]{mK} to minimize the heat capacity. In CUORE-0, this temperature is obtained using the same dilution refrigerator infrastructure as \qino~\cite{6}. The detector thermal link to the fridge is provided by the polytetrafluoroethylene supports which hold each crystal in the copper frames and by the golden wires used to carry the electrical signal.

The temperature of each crystal is continuously sensed by a neutron transmutation doped (NTD) thermistor~\cite{17} glued to the crystal surface. The thermistor converts the thermal signal to a voltage output which is digitized at an acquisition rate of \unit[125]{samples/s}. Signal pulses corresponding to thermal events are identified through a software trigger, and a \unit[5]{s} long window --- 1 s before and 4 s after the trigger --- is selected for further analysis. The initial second (pre-trigger) is used to establish the baseline temperature of the crystal just prior to the event. The pulse amplitude is used to determine the deposited energy. In order to monitor and correct for changes in detector gain due to temperature drifts,  a silicon resistor (heater)~\cite{26} is coupled to each crystal and is used to generate reference thermal signals every \unit[300]{s}~\cite{27,28}. A time-coincidence analysis can be performed to search for events that involve multiple crystals simultaneously. To account for the time response of the detector we use a coincidence window of $\pm$5~ms. As the measured event rate is approximately 1 mHz/crystal, the probability of accidental (i.e., causally unrelated) coincidences is extremely small ($\simeq 10^{-5}$).

In order to reduce background due to environmental radioactivity, the tower and cryogenic infrastructure are surrounded by several layers of shielding, including an internal low-background Roman lead layer and an external anti-radon box.
The details of the CUORE-0 detector design, operation and performance are described in~\cite{5,8}. The dilution refrigerator, shielding, and other cryostat components are those from the \qino experiment~\cite{6,7}.

\section{Analysis}
\label{sec:analysis}
In this work we exploit the multi-detector coincidence patterns expected to accompany $\beta\beta_{0^+}$ decays to maximize our sensitivity. The emitted electrons and gamma rays can interact in multiple crystals, producing a variety of experimental {\it{scenarios}} or signatures depending on the number of detectors involved in the process. We first identify all the possible signatures that can be detected and rank them by their expected sensitivity to $0\nu\beta\beta_{0^+}$ or $2\nu\beta\beta_{0^+}$ decays. Our determination of the most significant signatures to be used in the final analysis makes use of both real and simulated data.

\subsection{Signature identification}
\label{Sec:SigId}
To simplify the analysis we restrict ourselves to signatures in which the electrons are fully contained within the crystal where the decay took place. We further require that each individual de-excitation gamma is completely absorbed in a single crystal, thus discarding events where these gamma rays scatter but subsequently escape the crystal. Since the maximum number of gamma rays emitted in the decay is three (Figure \ref{fig:decay_scheme}), these choices imply that the maximum number of crystals involved in an event is four (one crystal contains the two electrons).  For every possible signature satisfying these conditions, both for the $2\nu\beta\beta_{0^+}$ and the $0\nu\beta\beta_{0^+}$ decay modes, we identify a single monochromatic line in one of the up-to-four crystals which can be used to estimate the decay rate through the fit described in section \ref{Sec:Fitting}.  For this reason, at least two crystals must be involved in $2\nu\beta\beta_{0^+}$ decay signatures, so that at least one of them records a monochromatic peak --- when only one crystal is involved the energy deposited is a continuous distribution spread over a 734 keV-wide range. 

The analysis therefore ultimately involves searching for the peak associated with each signature.  Considering all the possible gamma cascade patterns and the aforementioned constraints, a total of 57 signatures remain. However, only a few of them produce a sizable contribution to the half life sensitivity to double beta decay.  In the case of a peak search in the presence of nonzero background the half-life sensitivity has the following dependence on the experimental parameters: 
\begin{equation}
    T^{1/2}_{0^+}\propto\epsilon\sqrt{\frac{M\cdot t}{b\cdot\Delta E}},
\end{equation}
where $\epsilon$ is the total detection efficiency, $M\cdot t$ is the exposure, $b$ is the background rate per unit energy (background index) in the energy region of interest (ROI) and $\Delta E$ is the energy resolution near the ROI. While the exposure is always the same (35.2 kg$\cdot$y of TeO$_{2}$), the other three parameters vary significantly depending on the signature and associated peak under consideration. \\
We measure the energy resolution and its energy dependence directly from CUORE-0 data. We find the resolution exhibits a linear energy dependence which we parametrize as $\Delta E (E)=\Delta E(2615)\times (p_0+p_1\times E)$\cite{8}; here $\Delta E(2615)=4.9$ keV is the average FWHM of the 2615 keV line obtained during the calibration runs, $p_0=0.49\pm0.04$ and $p_1=(2.22\pm0.15)\times10^{-4}$. To avoid biasing our ranking procedure the background index for each signature is estimated from the CUORE-0 background model described in \cite{10} rather than using the CUORE-0 data directly. The model involves a full reconstruction of the measured spectra with highly detailed Monte Carlo simulations based on the \texttt{Geant4} package~\cite{19}.  The selection conditions (cuts) associated with each analysis signature are applied to the simulated data and the background index is evaluated in the relevant energy range. 
Finally, the efficiency term accounts for the probability that a $\beta\beta_{0^+}$ event is triggered, produces the multi-detector coincidence signature in question, and the peak is properly reconstructed at the expected amplitude.  The quantities involved in the efficiency calculation are described in detail in section \ref{Sec:Efficiency}.

\begin{table}
\begin{center}
\begin{tabular}{c|c|c|c|c|c}
\hline
Scen. & \multicolumn{3}{|c|}{Energy [keV]}            & $b$       & $\epsilon$ \\
\#    & Det.A         & Det.B         & Det.C         &  &            \\
\hline
1     & 734           & 536           & \textbf{1257} & 2.46E-6 & 6.47E-3    \\
2     & \textbf{1991} & 536           &               & 2.84E-5 & 1.47E-2    \\
3     & \textbf{734}  & 536           &               & 9.03E-4 & 2.71E-2    \\
4     & \textbf{1270} & 1257          &               & 1.22E-3 & 2.28E-2    \\
5     & 734           & \textbf{1257} &               & 3.52E-4 & 1.50E-2    \\

\hline		
\end{tabular}
\caption{The five most relevant scenarios that contribute to the total sensitivity to the neutrinoless decay channel by more than 1\%. $b$ is the background index (units: counts/keV/kg/y) and $\epsilon$ the total detection efficiency. The electrons are assumed to be always fully absorbed in the detector 'A'. The final fit is performed at the highest energy monochromatic peak measured in each signature, marked here in bold.
}
\label{tab:scen_0nu} 
\end{center}
\end{table}

\begin{table}
\begin{center}
\begin{tabular}{c|c|c|c|c|c}
\hline
Scen. & \multicolumn{3}{|c|}{Energy [keV]}             & b       & $\epsilon$ \\
\#    & Det.A         & Det.B         & Det.C         &  &            \\
\hline
1     & 0$\div$734     & 536           & \textbf{1257} & 6.28E-4 & 6.05E-3    \\
2     & 536$\div$734 & \textbf{1257} &               & 3.01E-2    & 2.25E-2    \\
3     & 734$\div$1270 & \textbf{1257} &               & 3.44E-2    & 1.18E-2    \\
4     & 1405$\div$1991 & \textbf{536}  &               & 5.91E-2    & 1.08E-2    \\
5     & 1320$\div$1405 & \textbf{536}  &               & 1.36E-2 & 4.11E-3    \\

\hline		
\end{tabular}
\caption{The five most relevant scenarios that contribute to the total sensitivity to the two neutrino decay channel by more than 1\%. $b$ is the background index (units: counts/keV/kg/y) and $\epsilon$ the total detection efficiency. The electrons are assumed to be always fully absorbed in the detector 'A'. The final fit is performed at the highest energy monochromatic peak measured in each signature, marked here in bold.
}
\label{tab:scen_2nu} 
\end{center}
\end{table}

We define our total sensitivity to $0\nu\beta\beta_{0^+}$/$2\nu\beta\beta_{2+}$ as the sum in quadrature of the sensitivities given by each signature. We consider a signature to be relevant if its contribution to the total sensitivity to the process exceeds 1$\%$. Only five such scenarios are identified for both the $0\nu\beta\beta_{0^+}$ and the $2\nu\beta\beta_{0^+}$ channels, and they are ranked by their sensitivity in Tables \ref{tab:scen_0nu} and \ref{tab:scen_2nu} respectively. The selected scenarios cover $\sim97\%$ and $\sim99\%$ of the total sensitivity for the $2\nu\beta\beta_{0+}$ and $0\nu\beta\beta_{0+}$ cases respectively.

\subsection{Data selection}
We first remove time periods where the data quality are poor; the effect of this is accounted for in the exposure. We next remove events that are either poorly reconstructed by our analysis or are non-signal-like using the pulse shape methods described in \cite{8}. We then impose cuts based on the deposited energy and on the event multiplicity (i.e., the number of crystals involved in the event). 
For the $0\nu\beta\beta_{0^+}$ case (Table \ref{tab:scen_0nu}), the energy and multiplicity rules a candidate event must pass are listed below.

\begin{itemize}
	\item In scenario 1, events must involve exactly three hits in the same coincidence time window. One of the three crystals must contain a signal with energy $E_{1}$ in the range $(734\pm 5\sigma_{734})$ keV. The notation $\sigma_{734}$ indicates the energy resolution at 734 keV, which is estimated from the resolution function reported in section \ref{Sec:SigId}. Another of the three crystals must have energy, $E_{2}$, in the $(536\pm 5\sigma_{536})$ keV range. No requirement is imposed on the energy deposited in the third crystal.
    \item In scenarios 2-5, events must have exactly two crystal hits. No requirement is set on the individual measured energies of the hits ($E_{1}$ and $E_{2}$), but rather on their sum, $E_{tot}=E_{1} + E_{2}$. Labeling as $E_{A}$ and $E_{B}$ the energies indicated in Table \ref{tab:scen_0nu} that are expected to be deposited in detector $A$ and $B$ respectively, and as $E_{AB}$ their sum ($E_{A}+E_{B}$), then $E_{tot}$ must lie in the range $(E_{AB}\pm 5\sigma_{AB})$, where $\sigma_{AB}=\sqrt{\sigma_{A}^2+\sigma_{B}^2}$. 
\end{itemize} 

We apply a similar logic to the $2\nu\beta\beta_{0^+}$ decay (Table \ref{tab:scen_2nu}), but the range defined by the continuous electron spectrum replaces the $\pm 5\sigma$ requirement. 

\begin{itemize}
	\item In scenario 1 events must involve exactly three hits. The energy of the first hit, $E_1$, is bound in the range $0<E_{1}<734$; the energy of the second hit, $E_2$, must be in the range $(536\pm 5\sigma_{536})$. No requirement is set for the hit on the third crystal.
    \item For scenarios 2-5, two hits are required in the same coincidence window. Again, we don't set a requirement on the individual hit energies, but on their sum, $E_{tot}=E_{1} + E_{2}$. The continuous electron spectrum, contained in detector A (Table \ref{tab:scen_2nu}), defines two energy limits, $E_{A}^{min}$ and $E_{A}^{max}$. We require that $E_{tot}$ satisfy the condition $E_{A}^{min}+E_{B} < E_{tot} < E_{A}^{max} + E_{B}$.
\end{itemize}

\subsection{Efficiency evaluation}
\label{Sec:Efficiency}

The detection efficiency is the probability that a $\beta\beta_{0^+}$ event is triggered and properly reconstructed, including the production of the required multi-detector coincidence signature. The detection efficiency is a product of several factors:

\begin{itemize}
	\item the probability that an event is triggered, which we estimate from the fraction of heater-induced events that are triggered and correctly\hyphenation{re-con-struc-ted} reconstructed at the expected amplitude;
    \item the probability to include only physical events whose pulse shape does not differ from the average behavior (pulse shape efficiency);
    \item the probability to correctly measure the number of crystals involved in an event (i.e., the event multiplicity);
    \item the fraction of $\beta\beta_{0^+}$ events that deposit energy according to a particular scenario.
\end{itemize}

The trigger and pulse shape efficiencies are derived in the same way described in \cite{8}. The multiplicity term is obtained differently depending on the number of crystals involved. The probability that an event involving a single crystal (multiplicity 1) is actually recorded as one is calculated using the ${}^{40}$K line at 1461 keV~\cite{8}: since it's the only $\gamma$ line emitted in the decay, the only way for it to be measured in a multiplicity$>1$ event is by chance. For events with higher multiplicity we don't have a way to evaluate the efficiency directly, so we resort to a statistical derivation. We take into consideration the two main effects that can alter the multiplicity of one event: accidental coincidences and pile-up. 

\begin{itemize}
	\item Accidental coincidences are events that happen simultaneously by chance and not due to causally correlated signals. This leads to an artificially increased multiplicity.
    \item Pile-up refers to the situation where two (or more) signals happen randomly on the same channel within the same 5 seconds-long signal window. Such events are marked as pile-up and ignored by the coincidence calculation. This artificially reduces the multiplicity; if, for example, two signals are coincident but one of them is removed due to pile-up, the second signal is recorded as a multiplicity one event.
\end{itemize}

The magnitude of these effects can be estimated with Poisson statistics, considering the measured average event rate of $1$ mHz/channel, a 5 second long signal window and a 10 ms coincidence window. The resulting efficiencies for events with multiplicities 1 through 4 are listed in Table \ref{tab:eff}.\\
The final term in the efficiency, related to $\beta\beta_{0^+}$ decay itself, is evaluated using dedicated Monte Carlo simulations. We use the same simulation software employed for the CUORE-0 background model to reproduce the effects of all decay patterns and compute the fraction of fully-contained $\beta\beta_{0^+}$ events for each of the 57 scenarios. Due to the high statistics of these simulations, the error associated to this efficiency term is extremely small ($<0.1\%$). 




\begin{table}
	\begin{center}
		\begin{tabular}{|l|c|c|}
		\hline
        Efficiency term & Efficiency $[\%]$ & Error $[\%]$\\
        \hline 
       	Trigger & 98.529 & 0.004\\
        Pulse shape & 93.7 & 0.7\\
        Multiplicity 1 & 99.6 & 0.1\\
        Multiplicity 2 & 99.2 & 0.1\\
        Multiplicity 3 & 98.8 & 0.2\\
        Multiplicity 4 & 98.4 & 0.2\\
		\hline		
		\end{tabular}
		\caption{Signal detection efficiency terms. Trigger and pulse shape efficiencies are common to all scenarios. Multiplicity 1 efficiency is calculated from ${}^{40}$K, while multiplicities 2-4 come from a statistical calculation. The same statistical calculation applied to multiplicity 1 events yields the same result ($99.6\%$).}
		\label{tab:eff} 
	\end{center}
\end{table}

\subsection{Fitting technique}
\label{Sec:Fitting}

\begin{figure*}
 \centering
 	\subfigure
   		{\includegraphics[width=5.5cm]{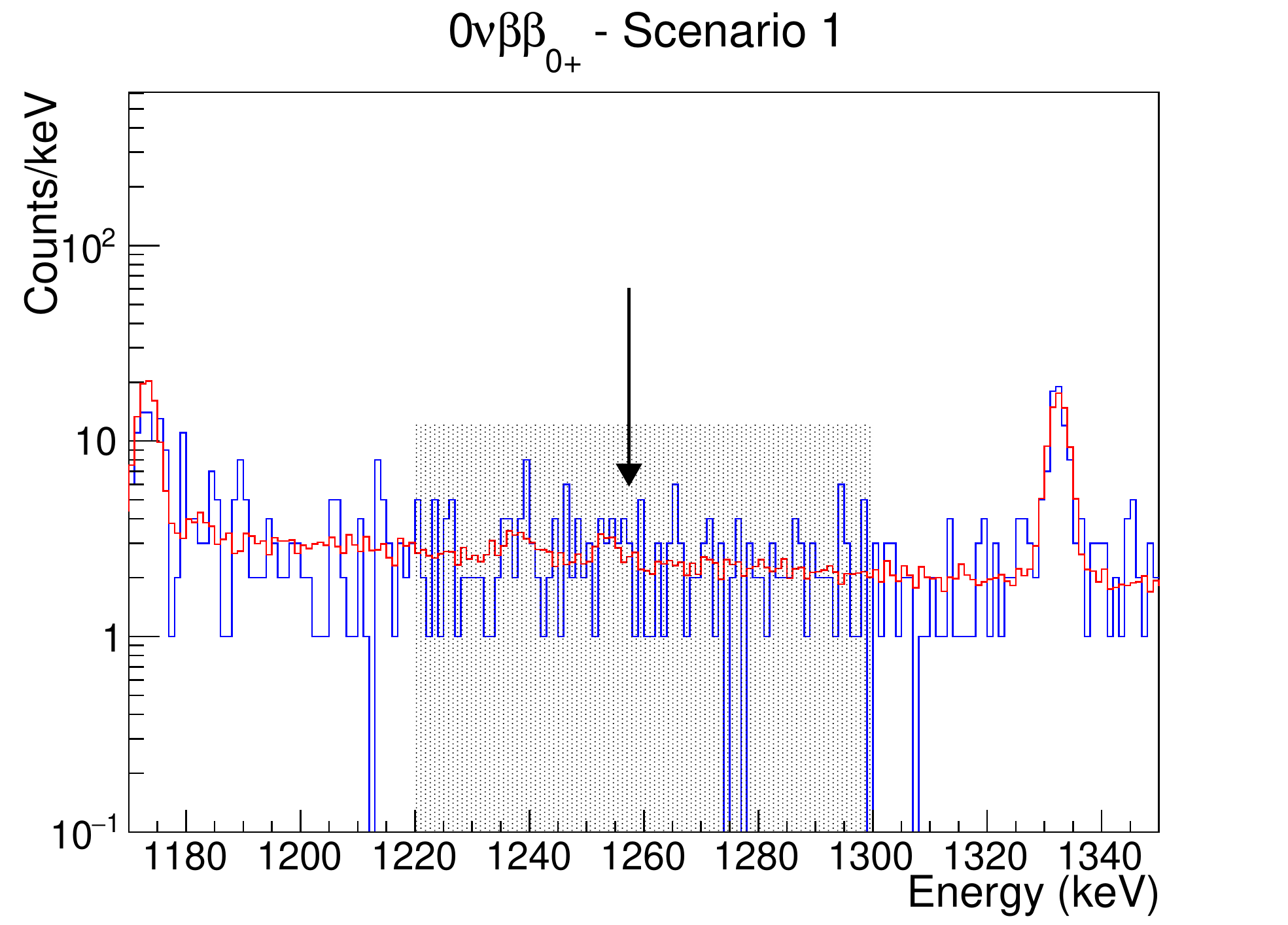}}
 	\subfigure
   		{\includegraphics[width=5.5cm]{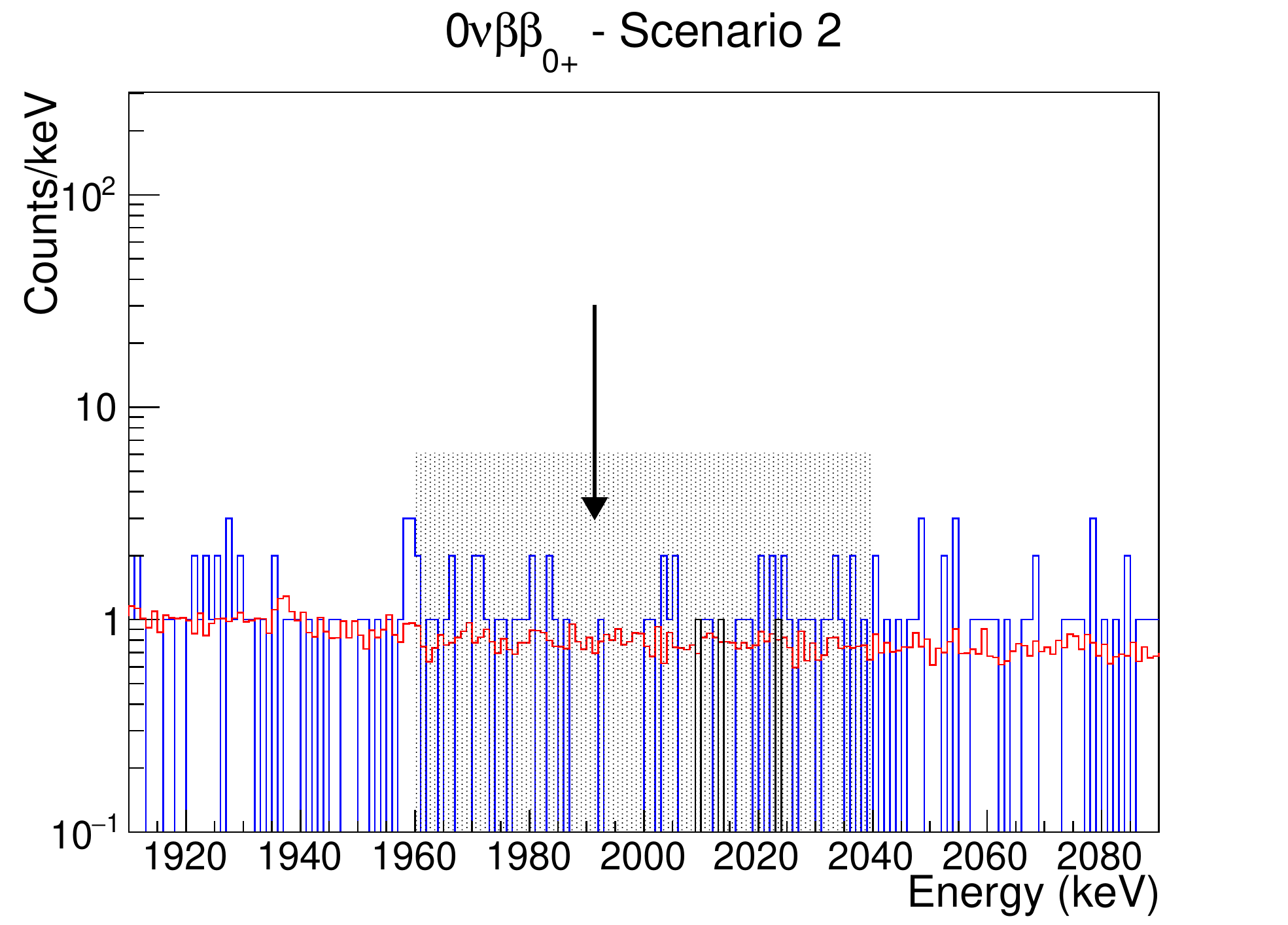}}
 	\subfigure
   		{\includegraphics[width=5.5cm]{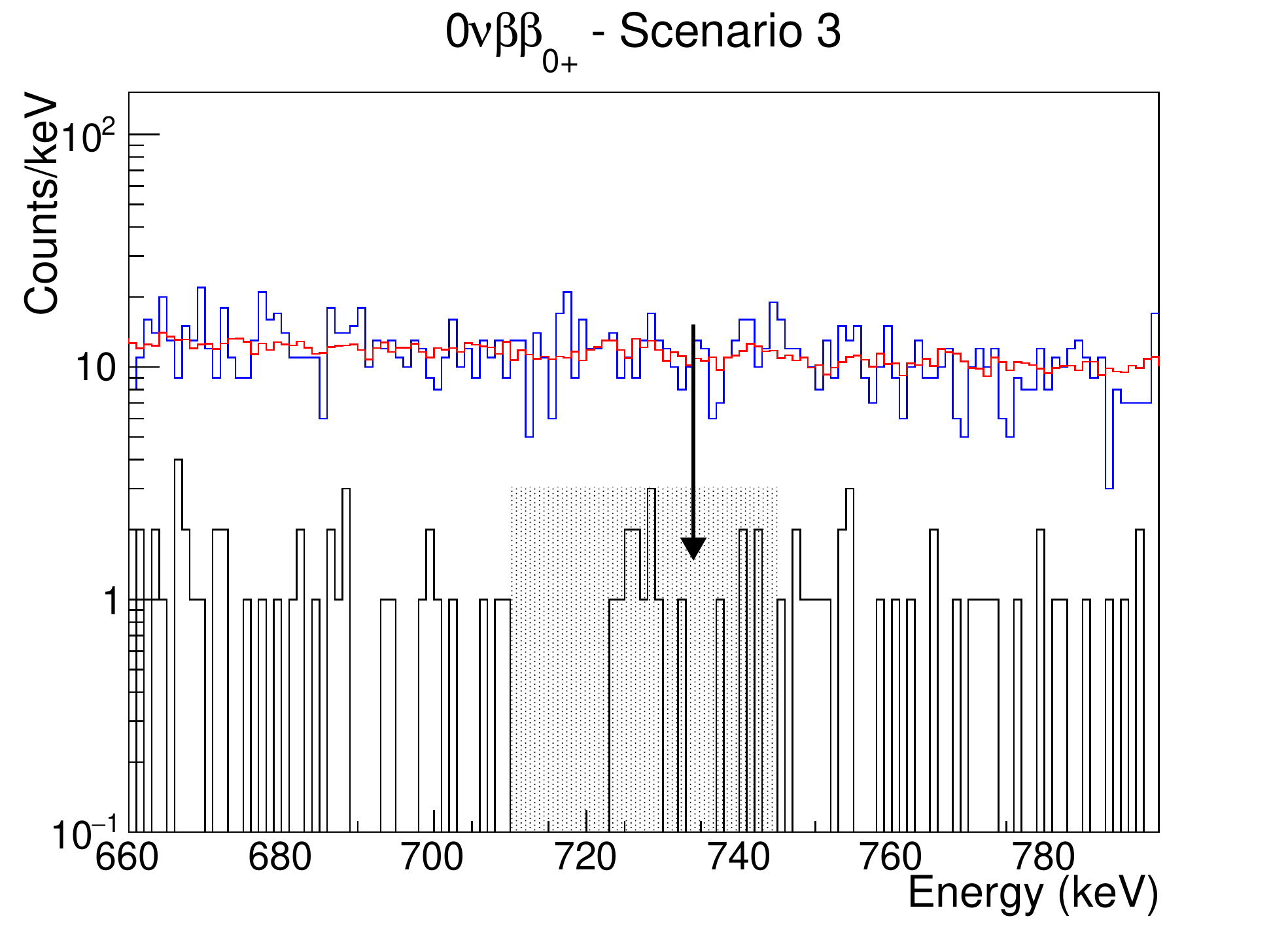}}
 	\subfigure
   		{\includegraphics[width=5.5cm]{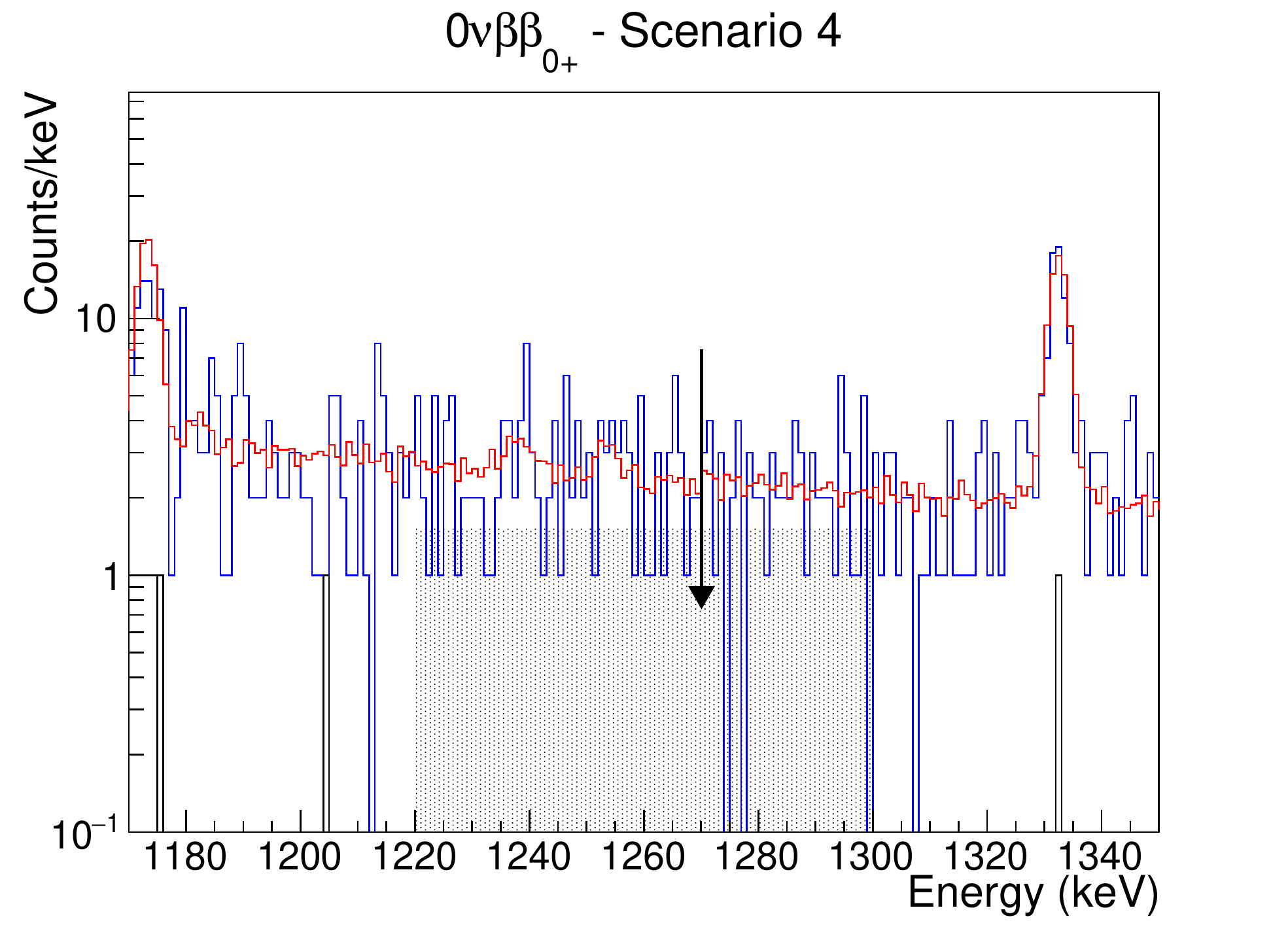}}
 	\subfigure
   		{\includegraphics[width=5.5cm]{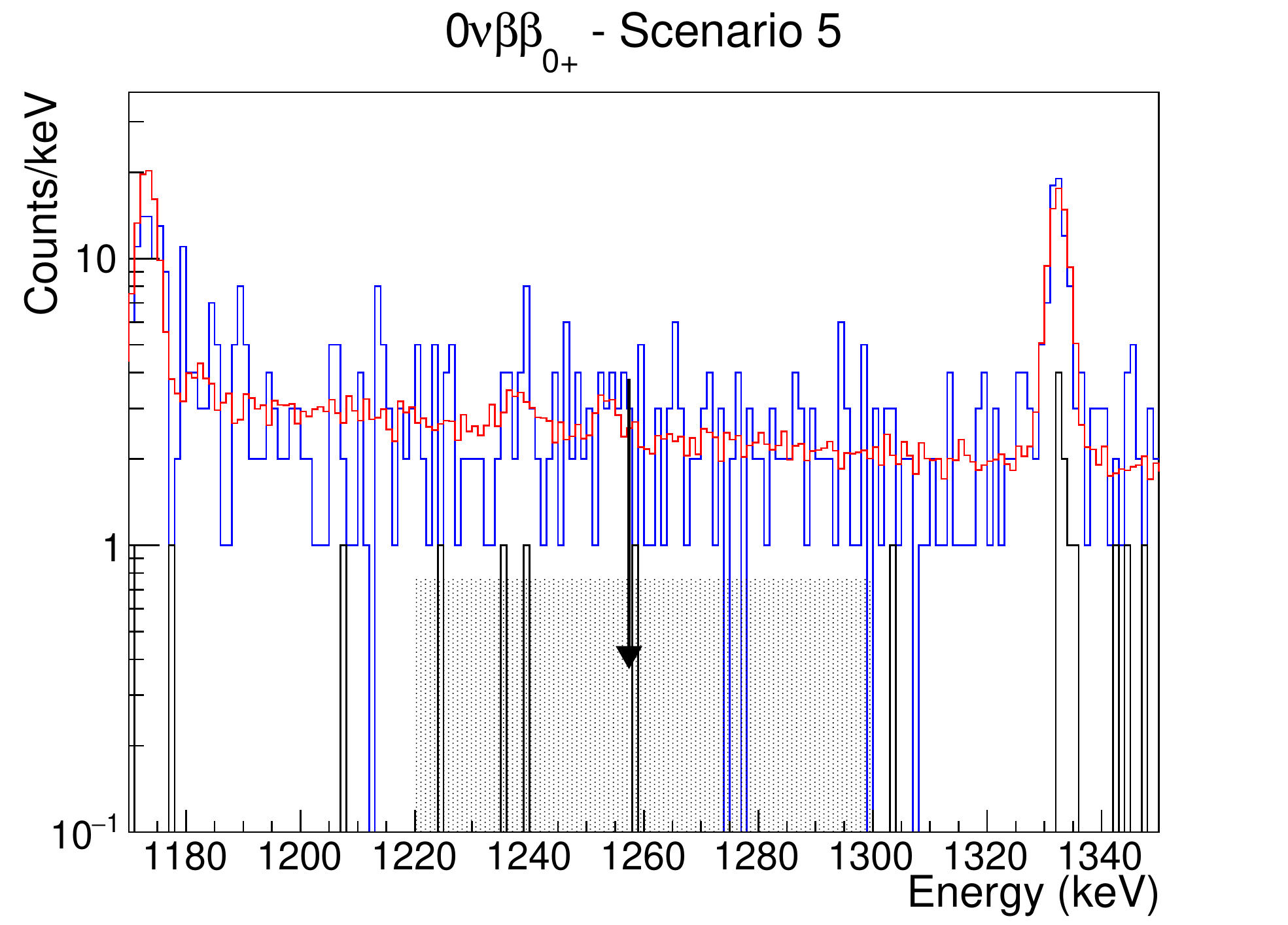}}
 	\caption{CUORE-0 spectra for the five signatures selected for the $0\nu\beta\beta_{0^+}$ decay. The blue histogram shows the data without any energy-related cut; the reconstruction of the background model is shown in red; in black are data with energy cuts. The box shows the fit range, while the black arrow points to the location of the expected $\beta\beta_{0^+}$ peak.}
    \label{fig:spectra_0nu}
\end{figure*}

\begin{figure*}
 \centering
 	\subfigure
   		{\includegraphics[width=5.5cm]{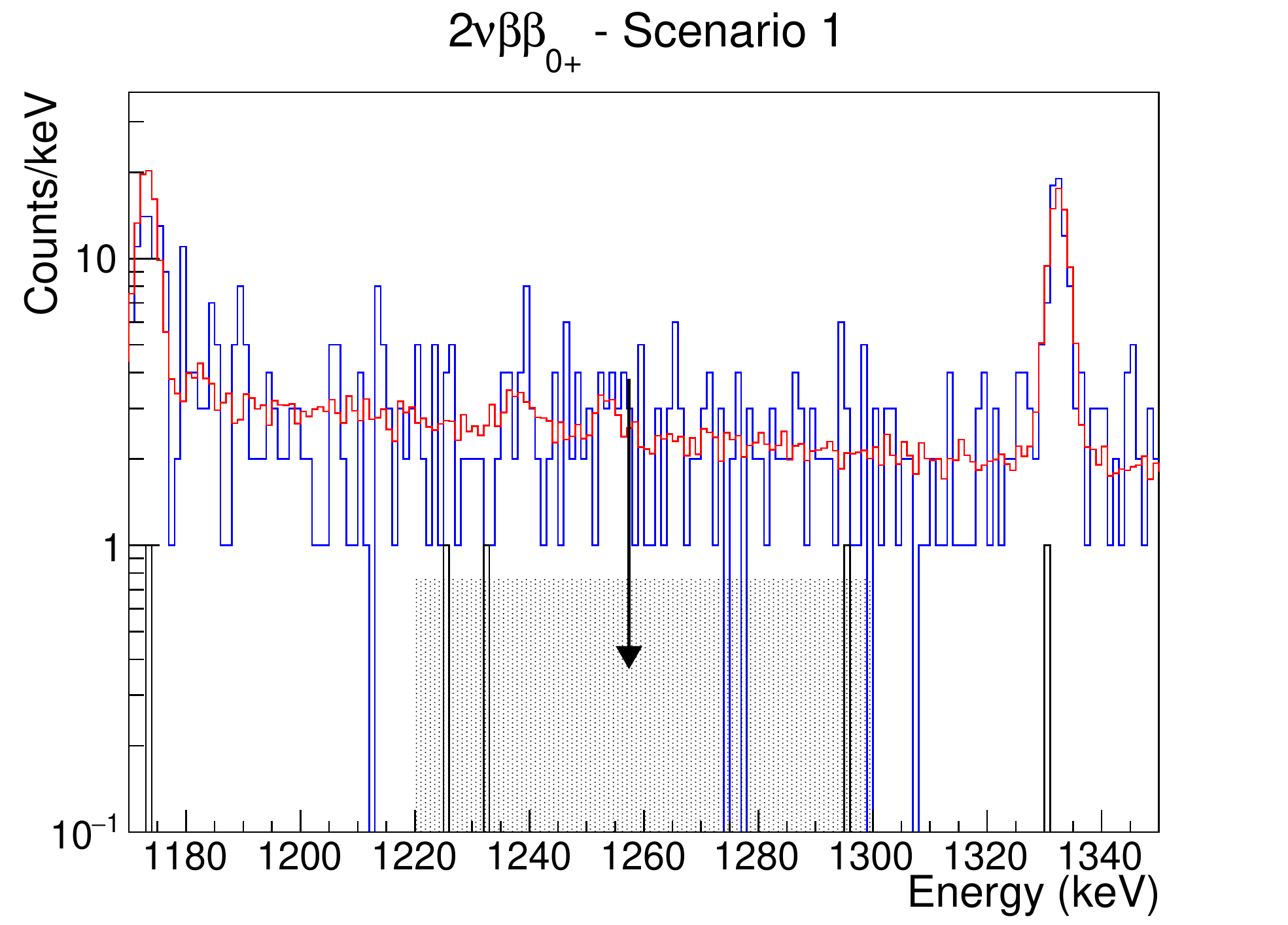}}
 	\subfigure
   		{\includegraphics[width=5.5cm]{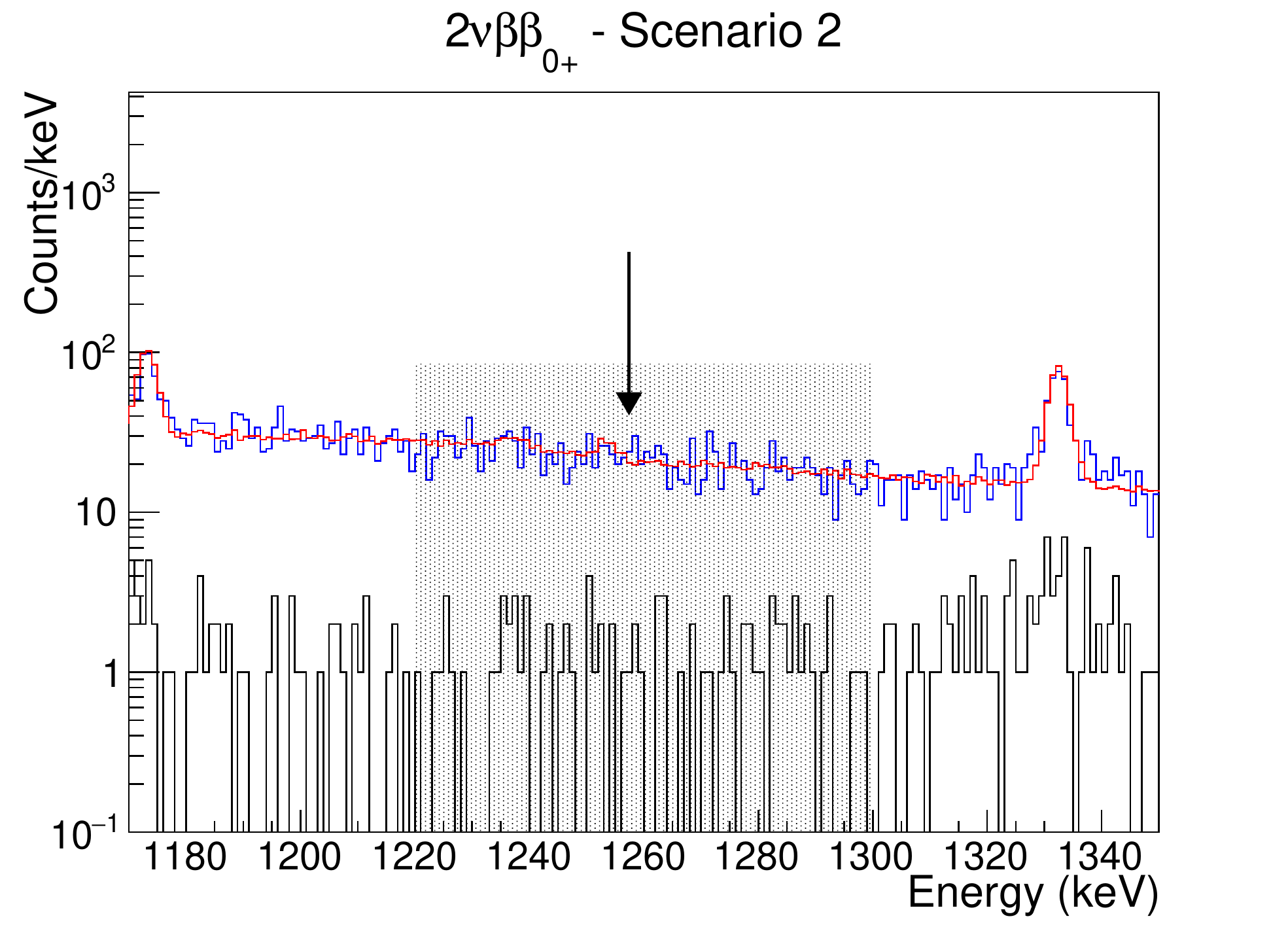}}
 	\subfigure
   		{\includegraphics[width=5.5cm]{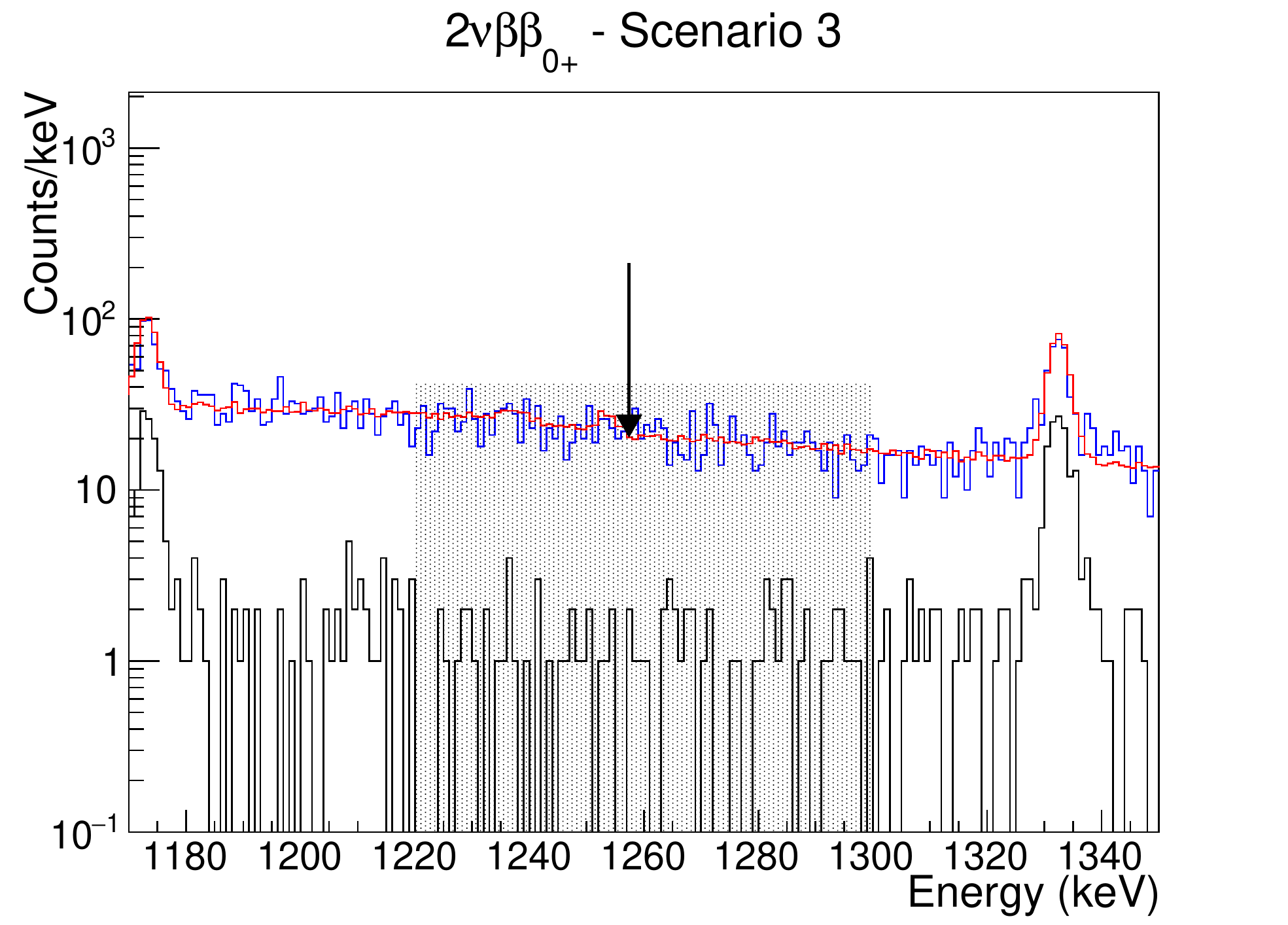}}
 	\subfigure
   		{\includegraphics[width=5.5cm]{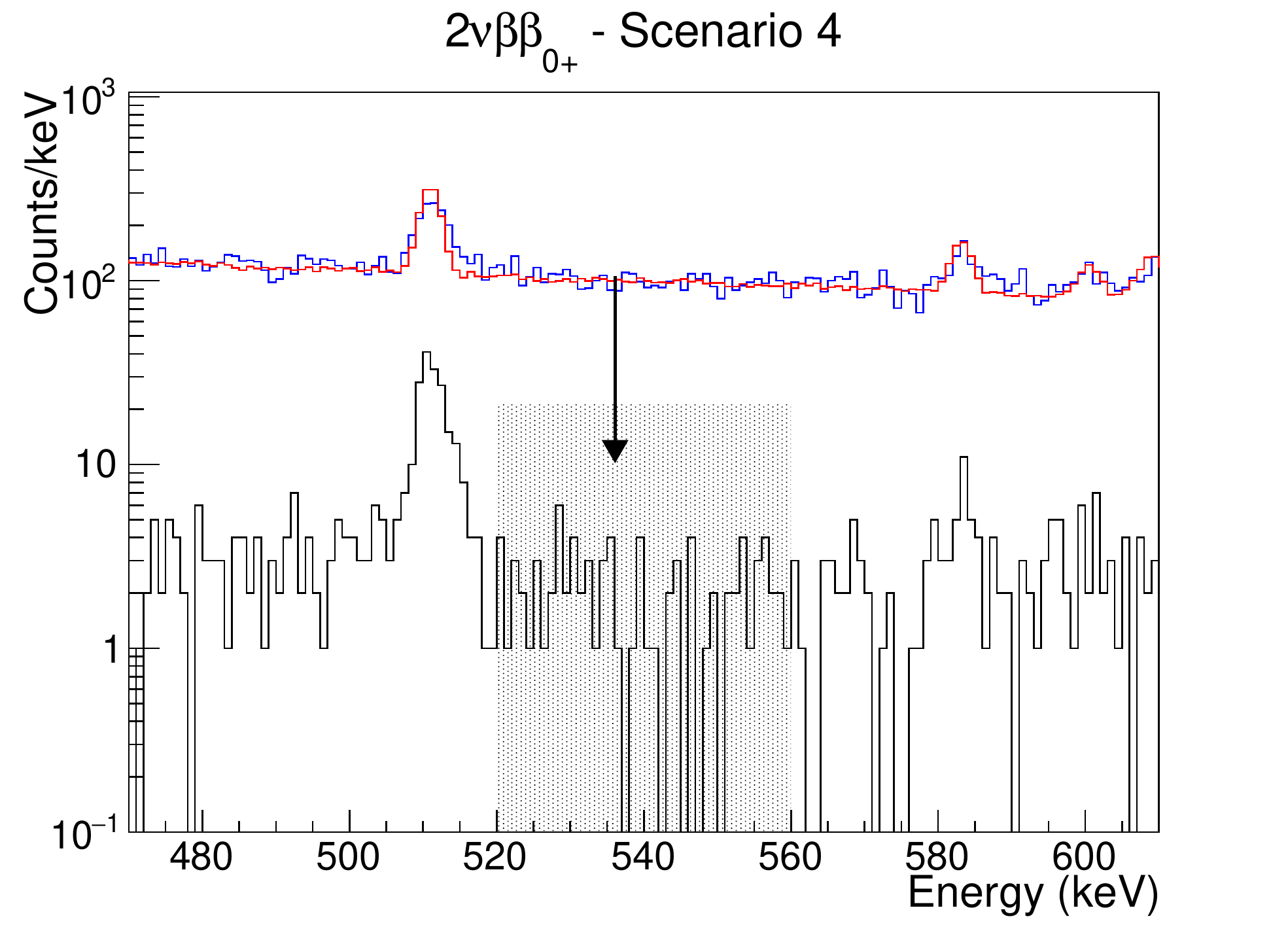}}
 	\subfigure
   		{\includegraphics[width=5.5cm]{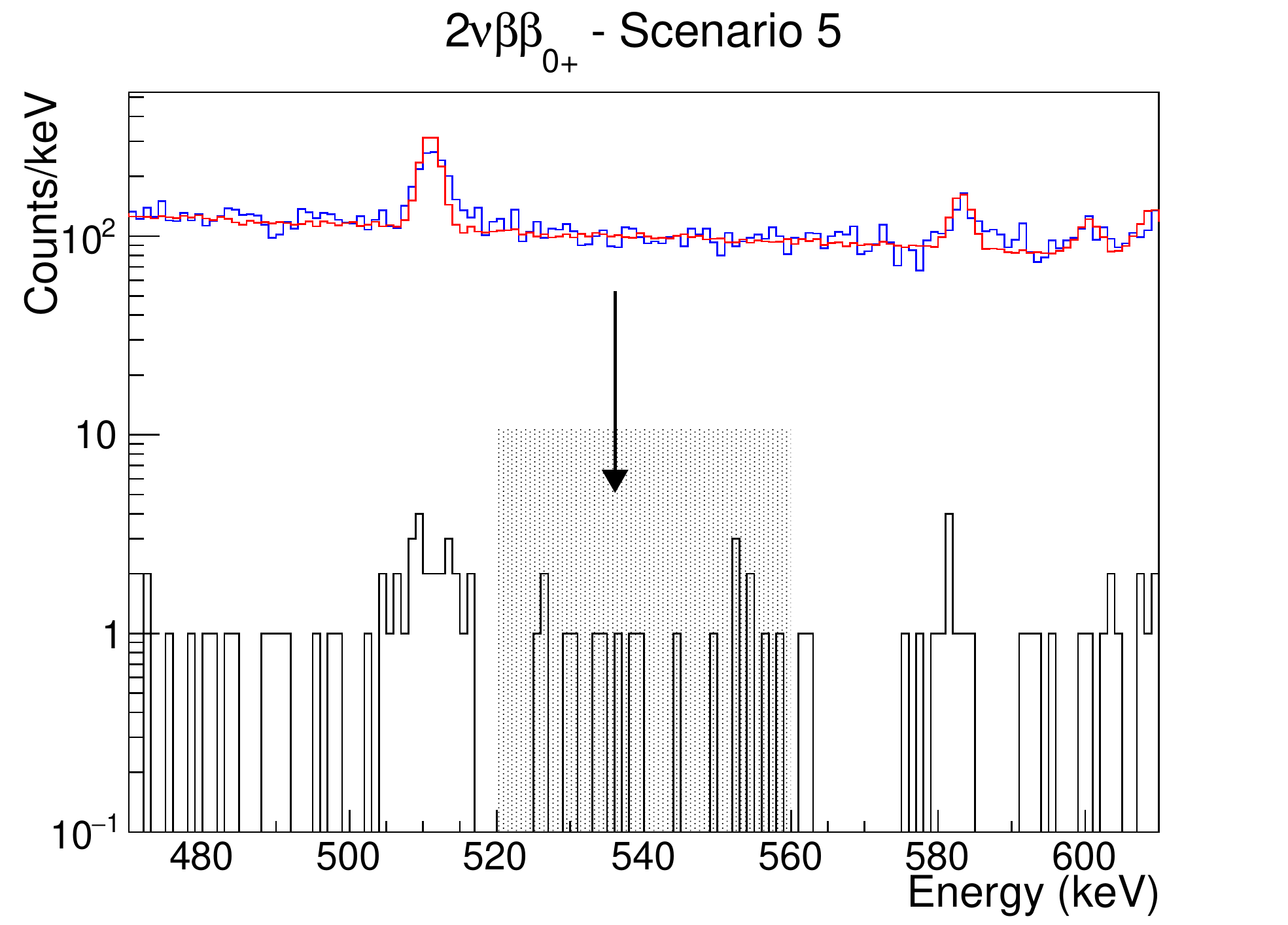}}
 	\caption{CUORE-0 spectra for the five signatures selected for the $2\nu\beta\beta_{0^+}$ decay. The blue histogram shows the data without any energy-related cut; the reconstruction of the background model is shown in red; in black are data with energy cuts. The box shows the fit range, while the black arrow points to the location of the expected $\beta\beta_{0^+}$ peak.}
    \label{fig:spectra_2nu}
\end{figure*}

The final step of the analysis procedure is to obtain for each scenario the energy spectrum selected for the fit. We choose to fit the spectrum of the crystal that records the monochromatic peak with the highest energy as, due to the shape of our observed spectra, it usually has the lowest background. To simplify the fits, ranges are chosen to exclude any peak from other $\gamma$ lines. This selection is based on the CUORE-0 background model\cite{10} rather than the data. The final spectra for the $0\nu\beta\beta_{0^+}$ decay search are shown in Figure~\ref{fig:spectra_0nu}, while those for $2\nu\beta\beta_{0^+}$ decay are shown in Figure~\ref{fig:spectra_2nu}. The strong background reduction achieved with energy-related cuts and the excellent agreement between real data and the background model are evident in these figures.

The fit function for each signature is the following:
\begin{equation}
	B_{const} + B_{lin}\cdot E + \dfrac{\epsilon\cdot t\cdot\Gamma_{\beta\beta}^{0^+} }{\sqrt{2\pi\sigma^2}}\cdot G(E_{\beta\beta}^{0^+},\sigma),
\end{equation}

where $B_{const}$ and $B_{lin}$ are parameters describing a linear background, $\epsilon$ is the detection efficiency, $t$ is the live time, $\Gamma_{\beta\beta}^{0^+}$ is the $\beta\beta_{0^+}$ decay rate, and $G$ is a gaussian function centered at $E_{\beta\beta}^{0^+}$ and with a resolution of $\sigma$. The expected values of $E_{\beta\beta}^{0^+}$ for each signature are indicated in bold in Tables \ref{tab:scen_0nu} and \ref{tab:scen_2nu}. We perform a simultaneous unbinned extended maximum likelihood (UEML) fit on the five signatures belonging to each $\beta\beta_{0^+}$ decay, using the RooFit fitting package \cite{11}. The fit parameters are constrained as follows:

\begin{itemize}
	\item the decay rate $\Gamma_{\beta\beta}^{0^+}$ is a common parameter for all five signatures;
    \item the two background components are independent for each signature;
    \item the detection efficiency is fixed at the value reported in Tables \ref{tab:scen_0nu} and \ref{tab:scen_2nu} for each signature;
    \item the exposure is fixed at $35.2$ kg$\cdot$y;
    \item the energy resolution is fixed at the value determined from the $\Delta E(E)$ curve reported in Section \ref{Sec:Efficiency};
    \item the peak position ($E_{\beta\beta}^{0^+}$) is fixed at the expected value reported in Tables \ref{tab:scen_0nu} and \ref{tab:scen_2nu} .
\end{itemize}

Each fit has 11 free parameters: two background parameters for each of the five signatures (for a total of 10) and the decay rate $\Gamma_{\beta\beta}^{0^+}$, which is common for all signatures.

\section{Results}
\label{sec:results}
We find no evidence of a $\beta\beta_{0^+}$ signal, either for the neutrinoless or the two neutrino decay mode. We set a $90\%$ confidence upper limit on the decay rates for the two processes: $\Gamma_{0^+}^{0\nu} < 8.8 \cdot 10^{-25} y^{-1}$, $\Gamma_{0^+}^{2\nu} < 2.8 \cdot 10^{-24} y^{-1}$. In turn, these correspond to the following lower limits for the half lives: 


\begin{center}
	$\tau_{0^+}^{0\nu} > 7.9 \cdot 10^{23}$ y, 90$\%$ C.L.,\\
    $\tau_{0^+}^{2\nu} > 2.4 \cdot 10^{23}$ y, 90$\%$ C.L.
\end{center}

We estimate the systematic uncertainties with a procedure identical to that applied in the analysis for $0\nu\beta\beta$ decay to the ground state~\cite{8,20}. We consider two components for the systematic uncertainty: one factor which scales with the decay rate ($\sigma_{scaling}$), and one which is independent of the decay rate ($\sigma_{add}$). We generate a large number of simulated spectra with a distribution taken from the best fit of each signature, but with the value of a single nuisance parameter modified by $1\sigma$; we then fit the simulated spectra with the unmodified parameter. To probe the value of $\sigma_{scaling}$, we repeat the analysis including in the simulated spectra a fake signal of variable strength. We regress the resulting best-fit decay rates against the simulated values to determine $\sigma_{add}$ and $\sigma_{scaling}$. This procedure is applied separately to get the systematic contributions from the uncertainty on efficiency, energy resolution and peak position. We also run the simulation without changing any parameter, to check for a possible fit bias. The resulting values for $\sigma_{add}$ and $\sigma_{scaling}$ are reported in Table \ref{Tab_syst_0nu} for $0\nu\beta\beta_{0^+}$ and in Table \ref{Tab_syst_2nu} for $2\nu\beta\beta_{0^+}$. In both cases, the dominant effect is a small negative bias.

\begin{table}
	\begin{tabular}{lcc}
		\hline\hline
    	& Additive ($10^{-24} y^{-1})$ & Scaling $(\%)$\\
   		\hline
    	Energy resolution & -0.05 & 0.17\\
        Peak position & 0.01 & 0.01\\
        Efficiency & 0.08 & 0.27\\
        Bias & -0.29 & 0.18\\
   		\hline\hline
	\end{tabular}
    \caption{Summary of systematic uncertainties on $\Gamma_{0^+}^{0\nu}$}
    \label{Tab_syst_0nu}
\end{table}

\begin{table}
	\begin{tabular}{lcc}
		\hline\hline
    	& Additive ($10^{-24} y^{-1})$ & Scaling $(\%)$\\
   		\hline
    	Energy resolution & -0.05 & 0.19\\
        Peak position & 0.01 & 0.02\\
        Efficiency & 0.12 & 0.31\\
        Bias & -0.35 & 0.24\\
   		\hline\hline
	\end{tabular}
    \caption{Summary of systematic uncertainties on $\Gamma_{0^+}^{2\nu}$}
    \label{Tab_syst_2nu}
\end{table}

\begin{figure}
	\begin{center}
		\includegraphics[width=\linewidth]{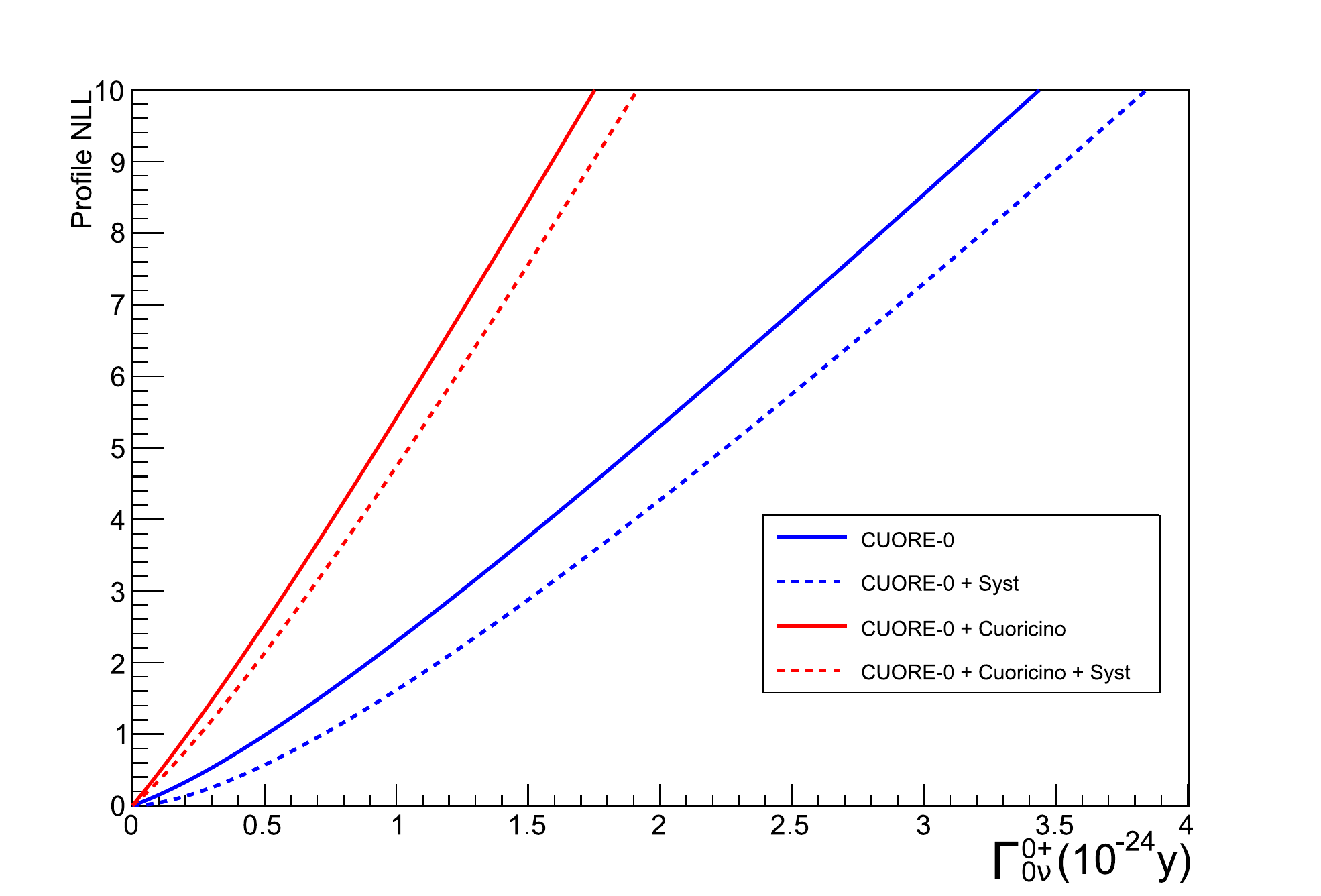} %
		\caption{Negative log likelihood (NLL) from CUORE-0 and the combination with CUORICINO for $0\nu\beta\beta_{0^+}$ decay}
		\label{fig:like0nu}
	\end{center}
\end{figure}

\begin{figure}
	\begin{center}
		\includegraphics[width=\linewidth]{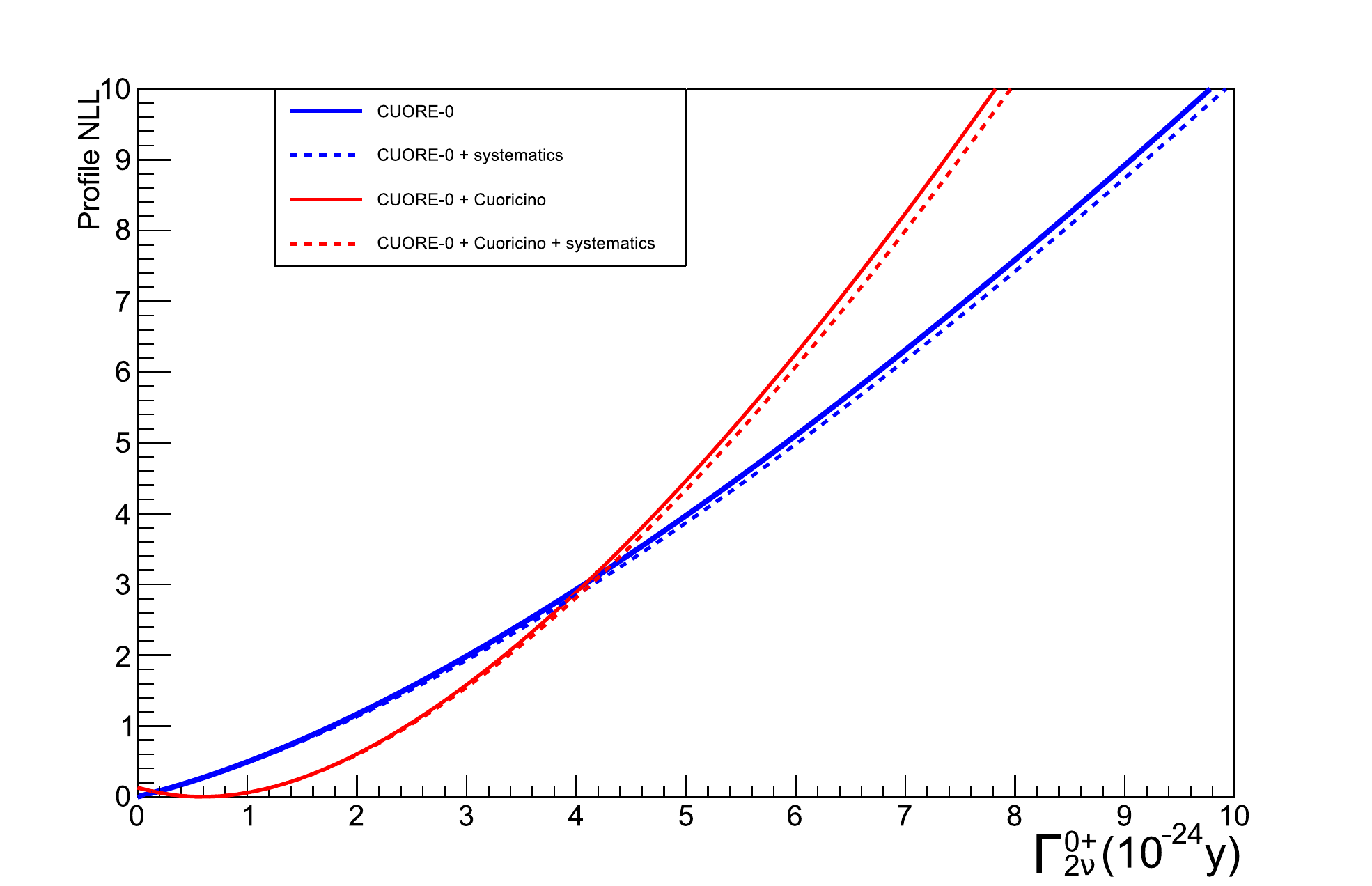} %
		\caption{Negative log likelihood (NLL) from CUORE-0 and the combination with CUORICINO for $2\nu\beta\beta_{0^+}$ decay}
		\label{fig:like2nu}
	\end{center}
\end{figure}

We combine the CUORE-0 likelihood curves with those from CUORICINO \cite{12} (Figure \ref{fig:like0nu} for $0\nu\beta\beta_{0^+}$ and Figure \ref{fig:like2nu} for $2\nu\beta\beta_{0^+}$). We set limits for the decay rates taking into account both the CUORE-0 systematic effects and the combination with the CUORICINO results: $\Gamma_{0^+}^{0\nu} < 4.8 \cdot 10^{-25} y^{-1}$, $\Gamma_{0^+}^{2\nu} < 2.7 \cdot 10^{-24} y^{-1}$. These yield the following limits on the half lives: 
\begin{center}
    $\tau_{0^+}^{0\nu} > 1.4 \cdot 10^{24} y$, 90$\%$ C.L.,\\
    $\tau_{0^+}^{2\nu} > 2.5 \cdot 10^{23} y$, 90$\%$ C.L.
\end{center}

Thanks to the improved background and analysis techniques, we achieve similar reach to CUORICINO with less than half of the exposure. The lower limits for the half lives of both the $0\nu\beta\beta_{0^+}$ and $2\nu\beta\beta_{0^+}$ decays obtained by the combination of the results from CUORE-0 and CUORICINO are the best currently available. CUORE will achieve even higher sensitivity, thanks to the improved background and to the powerful coincidence analysis made possible by its closely-packed 988 crystals.

\section{Acknowledgments}
The CUORE Collaboration thanks the directors and staff of the
Laboratori Nazionali del Gran Sasso and the technical staff of our
laboratories. This work was supported by the Istituto Nazionale di
Fisica Nucleare (INFN); the National Science
Foundation under Grant Nos. NSF-PHY-0605119, NSF-PHY-0500337,
NSF-PHY-0855314, NSF-PHY-0902171, NSF-PHY-0969852, NSF-PHY-1307204, NSF-PHY-1314881, NSF-PHY-1401832, and NSF-PHY-1404205; the Alfred
P. Sloan Foundation; the University of Wisconsin Foundation; and Yale
University. This material is also based upon work supported  
by the US Department of Energy (DOE) Office of Science under Contract Nos. DE-AC02-05CH11231,
DE-AC52-07NA27344, and DE-SC0012654; and by the DOE Office of Science, Office of Nuclear Physics under Contract Nos. DE-FG02-08ER41551 and DE-FG03-00ER41138.
This research used resources of the National Energy Research Scientific Computing Center (NERSC).

\bibliographystyle{spphys}       
\bibliography{bib}   

\end{document}